\documentclass[twocolumn, showpacs,amsmath,amssymb]{revtex4}
\usepackage{graphicx}
\usepackage{epstopdf}
\usepackage{bm}
\usepackage[latin1]{inputenc}
\usepackage{rotating}
\usepackage{longtable}
\usepackage{dcolumn}% Align table columns on decimal point
\usepackage{bm}% bold math
\usepackage{color}

\begin{document}

\preprint{APS/123-QED}

\title{Magic radio-frequency dressing of nuclear spins in high-accuracy optical clocks}

\author{Thomas Zanon-Willette$^{1,2}$\footnote{E-mail address: thomas.zanon@upmc.fr \\}, Emeric de Clercq$^{3}$, Ennio Arimondo$^{4}$}
\affiliation{$^{1}$UPMC Univ. Paris 06, UMR 7092, LPMAA, 4 place Jussieu, case 76, 75005 Paris, France}
\affiliation{$^{2}$CNRS, UMR 7092, LPMAA, 4 place Jussieu, case 76, 75005 Paris, France}
\affiliation{$^{3}$LNE-SYRTE, Observatoire de Paris, CNRS, UPMC, 61 avenue de l'Observatoire, 75014 Paris, France}
\affiliation{$^{4}$Dipartimento di Fisica "E. Fermi", Universit\`a di Pisa, Lgo. B. Pontecorvo 3, 56122 Pisa, Italy}
\date{\today}

\begin{abstract}
A Zeeman-insensitive optical clock atomic transition is engineered when nuclear spins are dressed by a non resonant radio-frequency field.  For fermionic species as $^{87}$Sr, $^{171}$Yb, and $^{199}$Hg, particular ratios between the radiofrequency driving amplitude and frequency lead to "magic" magnetic values where a net cancelation of the Zeeman clock shift and a complete reduction of first order magnetic variations are produced within a relative uncertainty below the $10^{-18}$ level. An Autler-Townes continued fraction describing a semi-classical radio-frequency  dressed spin is numerically computed and compared to an analytical quantum description including higher order magnetic field corrections to the dressed energies.
\end{abstract}


\maketitle

The interaction between magnetic fields and the nuclear/electronic magnetic moments represents a flexible tool for the control of the internal and external degrees of freedom in atoms or molecules, widely employed in precision measurements, frequency metrology, and coherent manipulations of quantum systems. In frequency metrology, the presence of magnetic fields may represent a limit on the realization of specific targets. In the attractive context of the "magic" wavelength combining a vanishing differential shift of the clock levels with the Lamb-Dicke regime greatly reducing the motional effects~\cite{YeKimble:2008,Katori:2011,Derevianko:2011}, the quest for "magic" magnetic field values where first order Zeeman shift and magnetic fluctuation of the atomic transition is annulled, was proposed in~\cite{Derevianko:2010} and studied experimentally in rubidium~\cite{Chicireanu:2011}. Best performances in optical clocks are accessible by using atomic transitions allowed by a weak hyperfine mixing mediated through a small spin-orbit coupling with a resolution at the mHz level.
To obtain even better performances, it has been proposed to use the bosonic even isotopes eliminating the nuclear spin and removing completely the first order Zeeman effect with a residual second order magnetic shift comparable to those of the ion standards~\cite{Taichenachev:2006}. However in order to get rid of strong cold collision frequency
shifts associated to the bosons, the present frequency metrology is concentrated on the fermionic species, as $^{87}$Sr~\cite{Ludlow:2008,Westergaard:2011,Yamaguchi:2011,Falke:2011}, $^{171}$Yb~\cite{Barber:2008,Lemke:2009}, $^{199}$Hg~\cite{Hachisu:2008,Yi:2011}. There, the first and second order Zeeman shifts contribute by one order
of magnitude above the projected 10$^{-18}$ fractional
uncertainty of the frequency standard~\cite{Ludlow:2008,Falke:2011,Yamaguchi:2011}.\\
\begin{figure}
\resizebox{8.0cm}{!}{\includegraphics[angle=0]{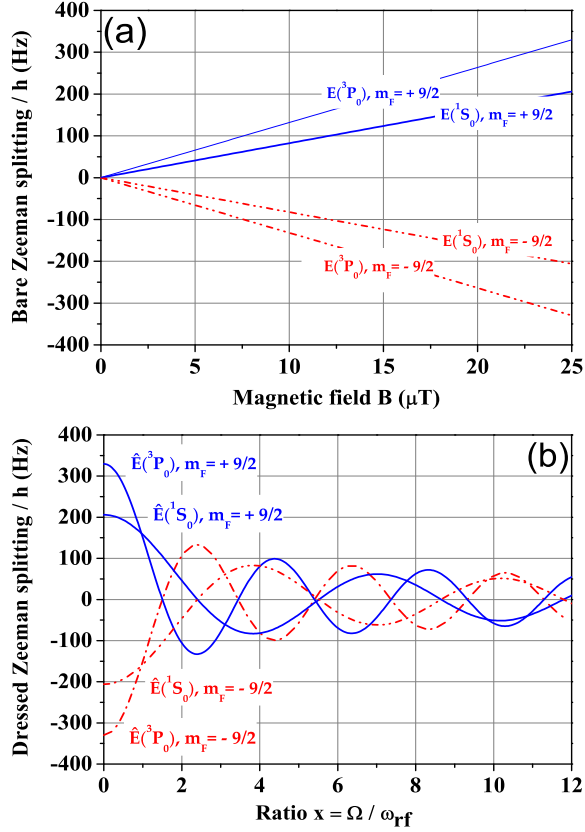}}
\caption{(Color online) (a) Bare Zeeman energy splitting $E_{m_F}$ of the $^{87}Sr$
$|^1S_0;m_F =\pm 9/2\rangle$ and $ |^3P_0;m_F = \pm 9/2\rangle$ clock states versus magnetic field $B$. (b) Dressed Zeeman energies $E^d_{m_F}$($^3$P$_0$,$^3$S$_0$) vs $x=\Omega/\omega_{rf}$ at $\omega/2\pi=2$ kHz and $B$=25 $\mu$T. The crossing nodes with a zero first order Zeeman shift determine the magic rf values.}
\label{Zeeman-shifts}
\end{figure}
\indent This work focuses on the atomic magnetic moment engineering with the target of getting rid of atomic properties sensitive to external electromagnetic fields.
A scheme based on the radiofrequency (rf) quantum engineering of fermionic atomic states is presented in order to produce levels experiencing a vanishing first-order Zeeman clock shift. The cancelation of first-order Zeeman shift applies also to the vectorial ac Stark shift, equivalent to an effective magnetic field, the only contribution of this kind  appearing in $^{171}$Yb and $^{199}$Hg. The basic idea of letting a paramagnetic system mimic a non-magnetic one originates from artificial or synthetic magnetism where an atomic Hamiltonian is created by proper electromagnetic fields  in order to simulate a given magnetic configuration~\cite{Dalibard:2011}.
Our work is inspired by the dressed-atom rf quantum engineering~\cite{CohenGueryOdelin:2011}, where the paramagnetic response for two species, atoms in~\cite{Haroche:1970} or atom/neutron in~\cite{Esler:2007}, is tuned into the resonance.\\
\indent The cancelation of the first order Zeeman effect is produced by the atomic dressing at a rf frequency much larger than the effective Larmor precession, equivalent to a frequency modulation of the nuclear magnetization and a shielded nuclear response to the static magnetic field. The different rf response for the ground and excited states of the clock transition leads to crossing  nodes in the energy diagram, where the atoms become non-magnetic. This  change from a paramagnetic system to an non-magnetic one shares a strong analogy with a Landau theory of phase transition. In addition at magic static field values the rf dressing  engineers a Zeeman-insensitive atomic clock. This magic cancelation arises from  the nonlinear magnetic Hamiltonian associated to the rf dressing of the two-electron system. Even if the dressing does not eliminate the second-order Zeeman contribution, its contribution to clock state separation is strongly decreased by an operation at a magic magnetic field.  The stability of the ratio between rf dressing amplitude and rf angular frequency required to produce a target non-magnetic state matched to the aimed optical clock accuracy is experimentally reachable. The present approach of a magic magnetic field cannot be applied to tensorial ac shifts.\\
\indent For the alkali-earth atoms, a modified Breit-Wills theory describes the action of a magnetic field $B$ producing linear and quadratic nuclear spin dependent Zeeman shifts for the doubly forbidden $|^{1}$S$_{0}\rangle \to |^{3}$P$_{0}\rangle$ optical clock transition~\cite{Boyd:2007}. For the $|F,m_F>$ Zeeman level the energy $E_{m_F}$ is
\begin{equation}
\begin{split}
E_{P,m_{F^{'}}}(^3P_0)&=m_{F^{'}}g_{P}\mu_{B}B+\Delta_B^{(2)}B^2,\\
E_{S, m_F}(^1S_0)&=m_{F}g_{S}\mu_{B}B,
\end{split}
\label{ZeemanShifts}
\end{equation}
where $m^\prime_F$ and $m_F$ are the upper/lower magnetic quantum numbers, $g_P$ and $g_S$ the Land\'e g-factors, with $g_P=g_S+\delta g$, and $\Delta^{(2)}_B$ the second order Zeeman contribution,  $m_F$ independent. We will focus our attention on fermionic systems spin polarized in the extreme Zeeman sub-levels~\cite{Boyd:2007,Baillard:2008} where a systematic average on the transition frequencies of optical transitions symmetrically placed around line center is currently applied to cancel the linear Zeeman
shift and to probe accurately the second order Zeeman correction. Their parameters are listed into the Supplemental Material. The Zeeman energies of the highest $|m_F|$ $^{87}$Sr clock levels vs $B$ are plotted in Fig. 1(a).\\
\indent We drive the clock atoms by a non-resonant rf field, linearly polarized and orthogonal to $B$, at angular frequency $\omega_{\rm rf}$ and (ground-state) Rabi frequency $\Omega$, see Supplemental Material. A strong modification of the Land\'e g-factor occurs in the regime where $\mu_B B \ll \hbar \omega_{rf}$. For dressing by a large number of rf photons, a perturbative quantum analysis  predicts a dressed Land\'e g-factor dependent on the zeroth-order Bessel function of the first kind $g^{\rm d}_j(x_j)= g_jJ_{0}(x_j)$, with $x_j=\Omega_j/\omega_{rf}$, $\Omega_j=g_j\Omega/g_S$, $(j=P,S)$~\cite{note}. That dependence was verified in experiments on atoms~\cite{Haroche:1970,HarocheCohen:1970,Esler:2007}, neutrons~\cite{Muskat:1987} and chromium Bose-Einstein condensate~\cite{Beaufils:2008}. It is valid for whatever spin value and equally spaced Zeeman levels~\cite{Series:1977}. When $\mu_BB \approx \hbar \omega_{rf} $, the $g^{\rm d}_j$ expression includes an additional B dependence given by~\cite{Hannaford:1973,Series:1977}
\begin{equation}
\begin{split}
g_{j}^{d}=g_{j}\left[J_{0}(x_j)-\left(\frac{g_{j}\mu_BB}{\hbar\omega_{rf}}\right)^2S(x_j)\right],
\end{split}
\label{approxgeff}
\end{equation}
$S(x)$ being a product of Bessel functions. However for the first two crossing nodes of the rf dressed Zeeman energies of Fig. 1(b) the following approximated analytical $S(x)$ expression given by~\cite{Ahmad:1974} provides the  required accuracy:
\begin{equation}
S(x)=\frac{16}{2025x^4}\left[\alpha(x)J_2(x)+\beta(x)J_4(x)-\gamma(x)J_6(x)\right];\\
\label{S-function}
\end{equation}
where functions are $\alpha(x)=75\left(5x^{2}-x^4/4\right),~\beta(x)=6\left(408-74x^{2}-23x^{4}/16\right),~\gamma(x)=145\left(3x^{2}-x^{4}/2\right)/49$.
When the dressed Land\'e g-factor of Eq.~\eqref{approxgeff} is substituted into the energies of Eq.~\eqref{ZeemanShifts}, the rf dressed energies contain both $B^2$ and $B^3$ nonlinear terms.\\
\begin{figure}
\resizebox{8.5cm}{!}{\includegraphics[angle=0]{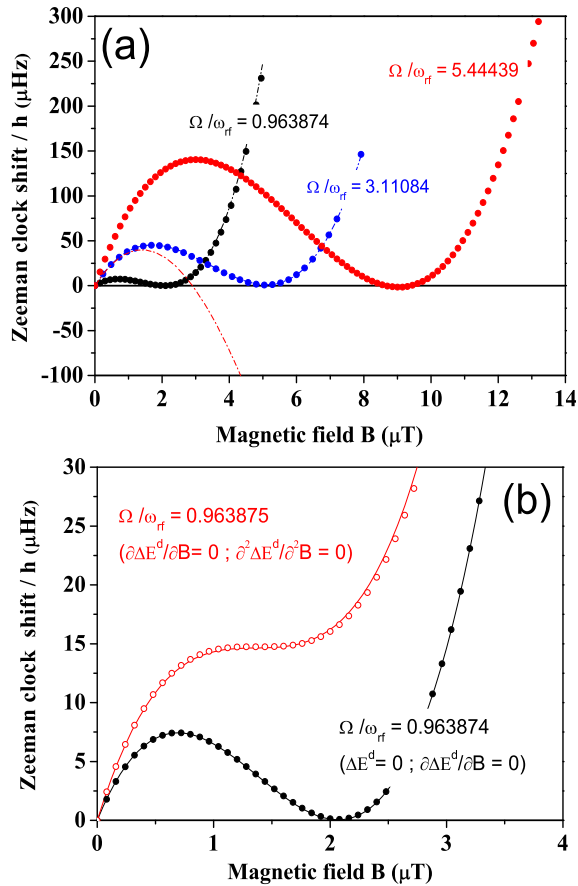}}
\caption{(Color online) $^{87}$Sr dressed Zeeman shifts of the  $\pi$ clock transition ($m_F=-9/2$  or $m_F=9/2$) vs $B$ for $\omega_{rf}/2 \pi =2~$kHz and the $\Omega/\omega_{rf}$ values at, or around, the crossing nodes in Fig.~\ref{Zeeman-shifts}. The curvature around the $B$ values of the shift minima is the dressed second order Zeeman shift. For the (b) open red data the $x$ is modified by one part in $10^6$ from the crossing value to a value where the second order Zeeman shift is annulled, in presence of  constant Zeeman bias. Open and closed dots $\bullet$ based on Eqs.~\eqref{dressed-solution} and \eqref{Autler-Townes-equation}. Lines based on Eqs.~\eqref{approxgeff} and \eqref{S-function} are approximated solutions providing a good description around the first two crossing  nodes only. 10 $\mu$Hz correspond to a $2.10^{-20}$ clock fractional shift.}
\label{Dressed-Zeeman-clock-shift}
\end{figure}
\indent We derive the exact rf dressed Zeeman energies $E^d_{m_{F}}(j)$ from the Autler-Townes continued fraction
\begin{equation}
E^d_{m_{F}}(j)=E_{m_{F}}(j)+m_{F}\hbar\frac{\Omega}{2}\frac{g_j}{g_S}\cdot L(j).
\label{dressed-solution}
\end{equation}
The function $L$, representing the $m=1/2$ dressed energy, normalized to the Rabi frequency~\cite{note}, for a  spin-1/2 system having $\omega_{21}$ energy splitting and rf dressed by a $\omega_{rf}$ field with Rabi frequency $\Omega$, is given by~\cite{Autler:1955}
\begin{equation}
\begin{split}
L=&\frac{1}{L+4\frac{\omega_{21}}{\Omega}\left(1-\frac{\omega_{\rm rf}}{\omega_{21}}\right)-\frac{1}{L-8\frac{\omega_{21}}{\Omega}\frac{\omega_{\rm rf}}{\omega_{21}}-\frac{1}{L+...}}}\\
&+\frac{1}{L+4\frac{\omega_{21}}{\Omega}\left(1+\frac{\omega_{\rm rf}}{\omega_{21}}\right)-\frac{1}{L+8\frac{\omega_{21}}{\Omega}\frac{\omega_{\rm rf}}{\omega_{21}}-\frac{1}{L+...}}}
\end{split}
\label{Autler-Townes-equation}
\end{equation}
The dressed nuclear Land\'e g-factor is obtained deriving the dressed energies with respect to B. Evaluation of dressed energies is done by retaining only a sufficient number of the quotients in each continued fraction needed to reach the desired accuracy. In practice nine quotients are necessary.\\
\begin{figure}
\resizebox{9cm}{!}{\includegraphics[angle=0]{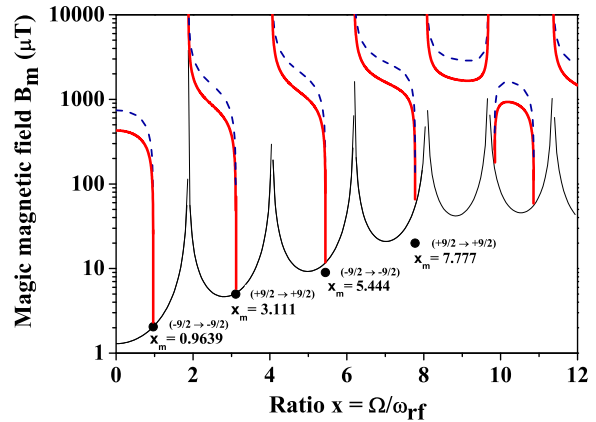}}
\caption{(Color online) $(B,x)$ branches, derived from Eqs.~\eqref{condition}, defining a zero clock Zeeman shift $ \Delta E^d = 0$ (thick solid red line) and a zero derivative $\partial\Delta E^d/\partial B=0$ (dashed blue line) for the alternating $m_F=-9/2$ and $m_F=9/2$ $\pi$ clock transitions in $^{87}$Sr. Their intersections determine the $(B_m,x_m)$ magic values also denoted by the solid dots $\bullet$ from a numerical evaluation of Eq.~\eqref{dressed-solution} and Eq.~\eqref{Autler-Townes-equation}. The thin solid black line dependence reports the $B(x)$ of Eq.~\eqref{magic-field}, but magic $B_m$ values occur only at magic $x_m$ parameters. All curves for $\omega_{rf}/2 \pi =2~$kHz.}
\label{dressing-mf-node}
\end{figure}
\indent The dressing field strongly modifies the Zeeman energies for each $m_{F}$ clock state as shown in Fig.~\ref{Zeeman-shifts}(b) for a given static field $B$. The energies follow mainly the zero-order Bessel function dependence. The different $^1S_0$ and $^3P_0$ sensitivity to the dressing created by $\delta g$ produces several crossings between the clock energies at particular $\Omega/\omega_{rf}$ ratios. The dressed Zeeman clock shift
\begin{equation}
\Delta E^d_{m_F \to m^\prime_F}= E^d_{m^\prime_F}(^3P_0)-E^d_{m_F}(^1S_0)
\end{equation}
is exactly compensated for specific rf dressing parameters, for instance in Fig.~\ref{Dressed-Zeeman-clock-shift}(a) at the $x=\Omega/\omega_{rf}\approx 0.96, 3.11, 5.44, ...$ for a $\pi$ transition. The $\omega_{rf}$ and $\Omega$ compensating values are determined by imposing the clock transition to be immune from the Zeeman shift, i.e., equal dressed magnetic energies.\\
\indent A more ambitious target is to derive the "magic" fields where the pure-Zeeman differential shift is zero and also  independent of the field value.  That applies to the dressed Zeeman clock shift plotted in Fig.~\ref{Dressed-Zeeman-clock-shift}(a) with  oscillations in the $B$ dependence. At $\omega_{rf}/2\pi=2$ kHz and $\Omega/\omega_{rf}=0.963874$ we have a magic magnetic field value $B_m\approx2.1$ $\mu$T which exactly cancels the full Zeeman shift for the $m_F=-9/2$, $\pi$ transition, experiencing in addition a  reduced $(B-B_m)^2$ sensitivity for the Zeeman energies. A  reduction by one-order of magnitude  for that second order sensitivity is obtained operating close to the higher order crossing nodes of Fig.~\ref{Zeeman-shifts}(b), the first two also shown in  Fig.~\ref{Dressed-Zeeman-clock-shift}(a) and corresponding to alternating $m_F=-9/2$ and $m_F=9/2$ Zeeman sub-levels. Fig.~\ref{Dressed-Zeeman-clock-shift}(b) shows that, by changing the $x$ parameter, a perfect cancelation of the $(B-B_m)^2$ magnetic sensitivity is reached, at expense of a clock constant bias.\\
\indent The "magic" $(B_m, x_m)$ values where the Zeeman shift and the first order sensitivity to weak field variations are simultaneously canceled are derived by imposing
\begin{equation}
\Delta E^d_{m_F\to m_{F'}}(x) = 0;\hspace{0.6cm}\left(\frac{\partial \Delta E^d_{m_F\to m_{F'}}}{\partial B}\right)_x=0,
\label{condition}
\end{equation}
The continued fraction solution of Eqs.~\eqref{dressed-solution} and \eqref{Autler-Townes-equation} determines the magic values associated to the above conditions. Fig.~\ref{dressing-mf-node} reports the graphical approach applied to derive these magic $B_m$ values. Notice that a suppression cannot be realized simultaneously for more than one $m_F$ spin dependent transition because  the $B$ square dependence of Eq.~\eqref{approxgeff} imposes for each optical transition a matched  dressed g-factor compensation.\\
\indent  As a good approximation, calculating the dressed energies through the effective Land\'e g-factor of Eq.~\eqref{approxgeff} we get the following expression of the magic $B_m$ field as a function of the $\omega_{rf}$ and $x_m$ parameters:
\begin{equation}
 B_m(\omega_{rf},x_m)=\frac{\hbar^3\omega_{rf}^2\Delta_B^{(2)}/(2\mu_B^3)}{m_{F^{'}}g_P^3S(\frac{g_{P}}{g_{S}}x_m)-m_Fg_S^3S(x_m)}
\label{magic-field}
\end{equation}
\indent  Table I reports at fixed $\omega_{rf}$ few magic pairs $(x_m,B_m)$, at increasing $B_m$ values, for $\pi$ polarization atomic clock of the fermionic species of present interest. The ~$\omega_{rf}/2\pi=2$ kHz choice, producing a consistent table for all the fermionic atoms, leads to very small magic field values for the mercury atom. Higher values, more easily manageable in the laboratory, are simply obtained by increasing the rf frequency and applying the Eq.~\eqref{magic-field} scaling. The accuracy of the dressed energy and the magic pairs strongly depends on the atomic parameters as tested by a numerical evaluation of the continued fraction. A complete "full-scale" calculations as reported in \cite{Derevianko:2010,Rosenbusch:2009} would be required for accurate magic numerical values including all the digits recommended for a correct evaluation. To highlight the resolution which should be targeted for canceling the Zeeman shifts below the $10^{-19}$ level, we have used a 6 digit resolution for figures when necessary.\\
\begin{table}
\caption{Magic $(x_{m},B_{m})$ pairs for $\pi$ and $\sigma^{\pm}$ $^{87}$Sr, $^{171}$Yb and $^{199}$Hg optical clock transitions based on Eqs.~\eqref{dressed-solution} and ~\eqref{Autler-Townes-equation}, at fixed $\omega_{rf}/2\pi=2~$kHz. The $x_m=\Omega_m/\omega_{rf}$ values are reported here with the experimental accuracy of the Landé factor, but a fractional shift below $10^{-19}$ requires a 6 digit resolution.}
\begin{tabular}{cccccc}
\hline
\hline
 &  & \textbf{$^{87}$Sr} &  & \\
$m_F \to m_{F'}$ & $-\frac{9}{2}\rightarrow-\frac{9}{2}$ & +$\frac{9}{2}\rightarrow+\frac{9}{2}$ & $-\frac{9}{2}\rightarrow-\frac{9}{2}$ & +$\frac{9}{2}\rightarrow+\frac{9}{2}$\\
$x_{m}$& 0.9639 & 3.111 & 5.444 & 7.777 \\
$B_{m}$ ($\mu$T) & 2.1 & 5.1 & 9.0 & 20.0 \\

\hline

 &  & \textbf{$^{171}$Yb} &  & \\

$m_F \to m_{F'}$ &$+\frac{1}{2}\rightarrow+\frac{1}{2}$ & $-\frac{1}{2}\rightarrow-\frac{1}{2}$ & $+\frac{1}{2}\rightarrow+\frac{1}{2}$ & $-\frac{1}{2}\rightarrow-\frac{1}{2}$ \\

$x_{m}$ & 0.9776 & 3.157 & 5.527 & 7.906 \\

$B_{m}$ ($\mu$T) & 0.08 & 0.12 & 0.33 & 0.68 \\

$m_F \to m_{F'}$ &$-\frac{1}{2}\rightarrow+\frac{1}{2}$ & $+\frac{1}{2}\rightarrow-\frac{1}{2}$ & $-\frac{1}{2}\rightarrow+\frac{1}{2}$ & $+\frac{1}{2}\rightarrow-\frac{1}{2}$  \\

$x_{m}$ & 1.826 & 4.107 & 5.543 & 6.954 \\

$B_{m}$ ($\mu$T) & 0.11 & 0.59 & 0.88 & 0.82 \\

\hline

 &  & \textbf{$^{199}$Hg} &  &\\

$m_F \to m_{F'}$ & $+\frac{1}{2}\rightarrow+\frac{1}{2}$ & $-\frac{1}{2}\rightarrow-\frac{1}{2}$ & $+\frac{1}{2}\rightarrow+\frac{1}{2}$ & $-\frac{1}{2}\rightarrow-\frac{1}{2}$ \\

$x_{m}$ & 0.9115 & 2.931 & 5.117 & 7.221 \\

$B_{m}$ ($\mu$T) & 0.02 & 0.05 & 0.11 &  0.34 \\

$m_F \to m_{F'}$ & $-\frac{1}{2}\rightarrow+\frac{1}{2}$ & $+\frac{1}{2}\rightarrow-\frac{1}{2}$ & $-\frac{1}{2}\rightarrow+\frac{1}{2}$ & $+\frac{1}{2}\rightarrow-\frac{1}{2}$  \\

$x_{m}$ & 1.674 & 3.599 & 4.566 & 6.388 \\

$B_{m}$ ($\mu$T) & 0.04 & 0.19 & 0.17 & 0.16 \\

\hline
\hline
\end{tabular}
\label{Magic-fields-table}
\end{table}
\begin{figure}
\includegraphics[width=8.5cm]{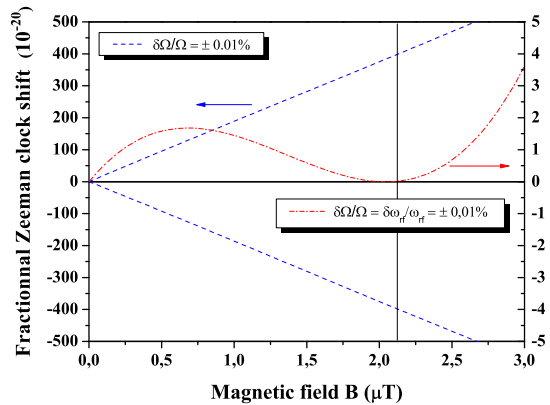}
\caption{(Color online) Fractional clock shift, measured in $10^{-20}$ relative units, for the $^{87}$ Sr  $m_F=-9/2$ $\pi$ transition, for a rf amplitude fluctuation by 0.01 percent (left scale) and for correlated variations of the same amount applied to both  amplitude  and rf frequency (right scale). Operating point $x_{rf}=0.96...$ and $\omega_{rf}/2\pi=2$ kHz, leading to the magic $B=2.1$ $\mu$T magic value.}
\label{Magnetic-sensitivity}
\end{figure}
\indent The magic field values are calculated at a fixed $\Omega/\omega_{rf}$ value, that implies a very large precision in the setting of the $\Omega$ and $\omega_{rf}$ parameters. For a practical application, the stability of those quantities becomes an important issue. While a very high stability of the rf frequency is not a problem, the $\Omega$ accuracy could be an issue. We have explored the $\Omega$ stability required in the operation of an optical clock, and the results are reported in Fig.~\ref{Magnetic-sensitivity}. A change in $\Omega$, or precisely a change in $x$ by one part in ten thousand corresponds to a fractional shift of the optical clock at the $10^{-17}$ level. In order to reach  the ultimate limit of the alkali-earths optical clocks~\cite{YeKimble:2008}, because only the $x$ ratio of the rf quantities is important for the rf engineering, the $\Omega$ variations may be compensated by acting on the rf frequency. Thus, a feedback on the rf frequency should reach the required ratio stability, in order to reach a $10^{-20}$ level in Fig.~\ref{Magnetic-sensitivity}. In practice the rf stability could be matched to the actual accuracy of the optical clock. For large $\omega_{rf}$ excursions, the efficiency of this compensation is limited by the $\omega_{rf}^2$ dependence in the $B_m$ numerator of Eq.~\eqref{magic-field}. \\
\indent We have verified that the shift produced by virtual transitions induced by the rf field between the $^3P$ fine structure levels is negligible compared to the finally aimed clock accuracy. The presence of  vectorial  ac shift contribution to the atomic energies   introduces an additional shift of the clock levels. When the ac  shift is comparable to the Zeeman shift,  the key features of the rf dressing, crossing nodes, compensations of the clock shift and magic values, are obtained also for this case, with "magic" $(x_m,B_m)$ pairs depending on the specific ac shift. Schemes for the rf compensation of the ac tensorial part appearing in the $^{87}$Sr optical clock only should be investigated.\\
\indent The combination of well-engineered optical-trapping potential and of rf quantum engineering represents an important tool in the investigation of alkali-earth clock systematics.  We have verified our scheme feasibility within the operation regime of the present optical clocks.  The averaging over the Zeeman components of the optical clock transition is directly performed by the rf dressing of the atomic system.
 For an implementation within an optical lattice where magnetic fields are created synthetically~\cite{Dalibard:2011}, the dressing magnetic field may be originated by the rf modulation of the lattice depth, at least for frequencies low enough for an atomic adiabatic following. \\
 \indent The synthetic rf controlled magnetism may be applied to other atomic/molecular and solid-state physics configurations.  Beside compensating the residual Zeeman contribution to a super stable optical clock based on a nuclear transition~\cite{Campbell:2012}, the rf engineering may be applied to design  an artificial quantum transition with specific Zeeman properties, as  a two-level superconducting system driven by an oscillatory field~\cite{Tuorila:2010}, and to the control of spin coherent dynamics and transport in semiconductor systems~\cite{Shen:2005,Ohe:2005}.  \\
\indent The authors  thank S. Bize, M. Glass-Maujean, A. Godone, C. Janssen, B. Laburthe-Tolra, R. Le Targat, A.D. Ludlow, and J.Ye for their inputs at different stages of this work. EA was supported by a MIUR PRIN-2009 grant.


\begin{thebibliography}{1}
\bibitem{note} Notice that owing to the different Land\'e g-factor in ground and excited states, the $^3$P$_0$ Rabi frequency is
$g_P\Omega/g_S$.
\end{thebibliography}
\end{document}

\title{Magic radio-frequency dressing of nuclear spins in high-accuracy optical clocks: Supplemental Material}

\author{Thomas Zanon-Willette$^{1,2}$\footnote{E-mail address: thomas.zanon@upmc.fr \\}, Emeric de Clercq$^{3}$, Ennio Arimondo$^{4}$}
\affiliation{$^{1}$UPMC Univ. Paris 06, UMR 7092, LPMAA, 4 place Jussieu, case 76, 75005 Paris, France}
\affiliation{$^{2}$CNRS, UMR 7092, LPMAA, 4 place Jussieu, case 76, 75005 Paris, France}
\affiliation{$^{3}$LNE-SYRTE, Observatoire de Paris, CNRS, UPMC, 61 avenue de l'Observatoire, 75014 Paris, France}
\affiliation{$^{4}$Dipartimento di Fisica "E. Fermi", Universit\`a di Pisa, Lgo. B. Pontecorvo 3, 56122 Pisa, Italy}
\date{\today}
\pacs{32.80.Ee,42.50.Ct,03.67.Lx}

\maketitle
The magic values for $B_m$ and $\Omega_m$ rely on the precise determination of the atomic parameters for the optical clock transition under investigation. In order to derive the values reported within the text of the present work, we have discovered that the published information on the required atomic parameters  is some time confusing. The main target of this Supplemental Material is to make clear some notations, and to list the numerical values of the Zeeman parameters used in the calculation in order to provide a correct comparison between the dressing of the different optical clocks.  In addition we precise our definition of the rf parameters in order to derive the precise value of the rf amplitude to be applied in the optical clock experiments.

\indent We define the Rabi angular frequency $\Omega_j$ as:
\begin{equation}
\Omega_j=\dfrac{g_j \mu_B B_{rf}}{\hbar},
\end{equation}
where $B_{rf}$ is the amplitude of the oscillating rf field, $g_j$ the Land\'e factor of the $j=(S,P)$ level, $\mu_B$ the Bohr magneton. This angular frequency is equivalent to the Larmor frequency $\omega_1$ defined in~\cite{Cohen:1973} and associated to the classical rf field amplitude. Our definition is four times larger than the frequency $b$ introduced in \cite{{Hannaford:1973},{Ahmad:1974}}. This explains why the coefficients of the function $S(x)$ of Eq. (3) in the main text are different of those of \cite{Ahmad:1974}.

The linear nuclear spin dependent Zeeman shift on $^{3}P_{0}$ and $^{1}S_{0}$ states depend on the Land\'e factors $g_S$ and $g_P$, see Eq. (1) of the main text. Their values  used in the numerical calculations are listed in Table \ref{atomic-parameters-table}. For the alkaline-earth-metal atoms, $^{87}$Sr and $^{171}$Yb, $g_S$ is calculated from the value of the shielded nuclear magnetic moment \cite{Olschewski:1972}. The $g_P$ values,  $g_P=g_S+\delta g$, are obtained from the experimental values of $\delta g$, for  $^{87}$Sr in~\cite{Boyd:2007}  and for $^{171}$Yb in~\cite{Lemke:2009} . The coefficient $\Delta\nu_{B}^{2}$ of the quadratic term was experimentaly measured for $^{87}$Sr in~\cite{Westergaard:2011}. That term has been measured on the fermion$^{171}$Yb~\cite{Lemke:2009}, but  we use the more precise boson $^{174}$Yb value of~\cite{Poli:2008}, since the quadratic Zeeman shifts of both isotopes should be identical, as measured for the boson~\cite{Baillard:2007} and fermion~\cite{Westergaard:2011} Sr isotopes. For $^{199}$Hg the $g_S$ factor is calculated from the value of the shielded nuclear moment ~\cite{{Cagnac:1961},{Gustavsson:1998}}, $g_P$ was measured in \cite{Lahaye:1975}, $\Delta\nu_{B}^{2}$  is given in~\cite{Hachisu:2008}. A recent publication \cite{McFerran:2012} gives experimental values in disagreement with these ones, but authors report that a factor 2 is probably missing (private communication). The precision of the atomic parameters is essential for a precise determination of the optical clock magic values while an experimental study of those values will lead to their better precision.\\
\begin{table}[h]
\caption{Atomic parameters of $^{1}S_{0}$ and $^{3}P_{0}$ fermionic states used in the main text}
\begin{tabular}[b]{cccc}
\hline
\hline
Atom  &  $^{87}$Sr &  $^{171}$Yb &  $^{199}$Hg \\
\hline
$I$  & $9/2$ & $1/2$ & $1/2$  \\
$g_S\mu_B/h$ (Hz/$\mu$T ) & 1.844 \cite{Olschewski:1972} & -7.439 \cite{Olschewski:1972} & -7.590 \cite{Cagnac:1961}  \\

$\delta g\mu_B/h$ (Hz/$\mu$T) & 1.084(4)~\cite{Boyd:2007}  & -4.20\cite{Lemke:2009}  & -5.712 \\

$g_P\mu_B/h$ (Hz/$\mu$T) & 2.928(4)  & -11.64  & -13.30 \cite{Lahaye:1975}  \\

$g_P/g_S$  ratio & 1.59 & 1.56 & 1.75 \\

$\Delta^{(2)}_{B}/h$ (Hz/mT$^2$)& -23.5(2)~\cite{Westergaard:2011}  & -6.1(1) \cite{Poli:2008} & -2.44~\cite{Hachisu:2008}  \\

\hline
\hline
\end{tabular}
\label{atomic-parameters-table}
\end{table}

\indent Note that the numerical values of Table I of the main text critically depend on the numerical values used for $g_S$ and $g_P$ (not $\delta g$ only), and for $\Delta^{(2)}_{B}$. The comparison between the strontium, ytterbium and mercury parameters evidences that the ratio of the nuclear magnetons between excited and ground states is similar for the three species. Instead a large difference is associated to the second order Zeeman shift, being larger in strontium by a factor between four and ten. Such difference leads to the magic magnetic field values reported in the Table I of the main text at fixed $\omega_{rf}$  very low for ytterbium and mercury.  For an implementation of the dressing of those atoms, it may be convenient to increase  $\omega_{rf}$ and bring the magnetic fields in a range where a better control is reached.